\documentclass[11pt]{article}
\usepackage{epsfig}
\usepackage[usenames]{color}
\usepackage{amsmath}
\usepackage{amsfonts}
\usepackage{amssymb}
\usepackage{mathrsfs}
\newtheorem{proposition}{Proposition}[section]
\newcommand{\bpr}{\begin{proposition}}
\newcommand{\epr}{\end{proposition}}

\newcounter{Roman}
\addtocounter{Roman}{1}

\newcommand{\beq}{\begin{equation}}
\newcommand{\eeq}{\end{equation}}
\newcommand{\bea}{\begin{eqnarray}}
\newcommand{\eea}{\end{eqnarray}}
\newcounter{saveeqn}

\DeclareMathOperator*{\esssup}{ess \, sup}

\newcommand{\s}{\scriptstyle}

\newcommand{\bfw}{{\bf w}}
\newcommand{\bfx}{{\bf x}}
\newcommand{\bfy}{{\bf y}}
\newcommand{\bfz}{{\bf z}}

\newcommand{\sign}{{\rm sign}}

\newcommand{\tlA}{\tilde{A}}
\newcommand{\tlH}{\tilde{H}}
\newcommand{\tlD}{\tilde{D}}
\newcommand{\tlDl}{\tilde{\Delta}}
\newcommand{\tlpsi}{\tilde{\psi}}

\newcommand{\tlphi}{\tilde{\phi}}

\input epsf
\begin{document}  

\begin{center}{\Large\bf  Nonlocal and quasi-local field theories}\\[2cm] 
{E. T. Tomboulis\footnote{\sf e-mail: tomboulis@physics.ucla.edu}
}\\
{\em Department of Physics and Astronomy, UCLA, Los Angeles, 
CA 90095-1547} 
\end{center}
\vspace{1cm}

\begin{center}{\Large\bf Abstract}\end{center}
We investigate nonlocal field theories, a subject that  has attracted some renewed interest in connection with nonlocal gravity models. We study, in particular, 
scalar theories of interacting delocalized fields, the delocalization being 
specified by nonlocal integral kernels. We distinguish between strictly nonlocal and quasi-local (compact support) kernels and impose conditions on them  to insure UV finiteness and unitarity of amplitudes. We study the classical 
initial value problem for the partial integro-differential equations of motion in detail. We give rigorous proofs of the existence but accompanying loss of uniqueness of 
solutions due to the presence of future, as well as past, ``delays", a manifestation of acausality. 
In the quantum theory we derive a generalization of the Bogoliubov 
causality condition equation for amplitudes, which explicitly exhibits the corrections due to nonlocality. 
One finds that, remarkably, for quasi-local kernels all acausal effects are confined within the compact support regions.  We briefly discuss the extension to other types of fields and prospects of such theories.

\vfill
\pagebreak

\section{Introduction} 
\setcounter{equation}{0}
\setcounter{Roman}{0}

Nonlocal field theories is a subject with long, albeit spotty, history. Despite the success of perturbative renormalization in QED in the late forties, the idea that local interactions may be a low energy approximation to fundamental underlying nonlocality of interactions continued to be prominent in the fifties and the subject of many investigations \cite{F}.  
Subsequently, nonlocality was considered mostly in the context of axiomatic field theory \cite{Ef}. In more recent years it has attracted renewed interest in connection with nonlocal theories of gravity 
\cite{TT} - \cite{Frol}, as well as the nonlocality of string field theory vertices and various nonlocal models in cosmology and other areas, see \cite{BK} and extensive reference list  therein.   

Despite this past work, basic issues have been left in a murky state. 
It has long been realized, more or less explicitly, that UV finiteness (or at least superrenormalizability in the presence of gauge interactions) can be achieved by nonlocal interactions. 
At the same time, unitarity can be preserved, at least perturbatively, provided appropriate analyticity conditions can be imposed on the nonlocal interactions. Causality, however, is a central concern whose investigation has remained woefully inadequate, both in the classical theory, where it is inexorably connected with 
the mathematically proper formulation of the initial value problem (IVP), and in the  
quantum theory. 

In this paper we address some of these issues in a more systematic way in the simplest 
field theory context, i.e. scalar field theories. (We comment on the extension to other theories 
in the last section.) To set the stage we recall  
that often, both in the older and more recent literature, nonlocal interactions have been introduced through the insertion of an operator of the general form 
\beq 
\hat{{\cal F}}(x-y) = e^{f(\ell^2 \partial_x^2)} \delta^4(x-y) \, , \label{deloc-a}
\eeq
where, in many instances, $f(z)$ is an entire analytic function. Note that $e^f$ is then itself an entire function possessing 
no zeroes anywhere in the complex plane. In many cases $f(z)$ is simply a polynomial. 
For example, $f(z) =z$ gives a generic form of vertices in string field theory, i.e., vertices of the form 
$[ (\exp \ell^2 \partial_x^2) \phi(x)]^3$ for, say, the dilaton field $\phi$. 
For Euclidean signature this has the explicit integral kernel form 
\beq 
 e^{\, \ell^2 \partial_x^2} \,\phi(x) =\left(\frac{\sqrt{\pi}}{\ell}\right)^d  \int d^dy\,  e^{-{1\over 4 \ell^2} (x-y)^2} 
\phi(y) \equiv \int d^dy \, F(x-y) \phi(y) \, . 
\label{deloc-b}
\eeq
For Minkowski signature, though, the operator on the l.h.s. of (\ref{deloc-b}) is ill-defined (except, of course, at $d=1$).   
This is the reason why only  even power polynomials for $f$ were  considered in the older non-local QFT literature,\footnote{In the fifties (\ref{deloc-c}) was referred to as the ``decay" choice as opposed to the  ``oscillatory" choice involving rational functions. \label{F2}} e.g., 
\beq 
e^{ -(\ell^2 \partial_x^2 )^2} \,\phi(x) = \int d^4 y \, F(x-y) \phi(y) \, ,
\label{deloc-c}
\eeq
where 
\beq 
F(x-y) = \left({\sqrt{\pi}\over \ell}\right)^4 \int d^4k \, e^{- (\ell^2 k^2 )^2}\,  e^{-ik\cdot(x-y)}   \,.     \label{deloc-d}
\eeq
This ensures well-defined kernels on the whole real axis $-\infty < k^2 < \infty$. 
In the more recent literature more elaborate choices of well-defined kernels 
employing transcendental entire functions $f(z)$ have been considered \cite{TT}, \cite{Mod1}.

For our purposes, there are three features of the nonlocal interactions introduced via 
(\ref{deloc-a})  
that are of interest. First, $f$ is such that the Fourier transform kernel $\hat{F}(k^2)$ is an entire analytic function of $k^2$. This is connected to unitarity in amplitudes.  
The second feature is that the resulting interactions, even though they may exhibit rapid or exponential fall-off, are truly non-local: any two spacetime points $x$ and $y$ are connected by the interaction integral kernels $F(x-y)$.  This should generally imply some acausal behavior.

Finally, since entire functions possess convergent series expansions\footnote{One should keep in mind, however, that any truncation of such an expansion, at any order, will fail to reproduce all the remarkable properties of transcendental entire functions, such as, for example, those embodied in Piccard's theorems. \label{FPicc}}  about any point, (\ref{deloc-a}) may be viewed as representing ``infinite order" derivative interactions. 
Attempts have been made historically to deal with such interactions in some sort of Hamiltonian formalism with infinite order derivatives. This has apparently been the source of a great deal of confusion in the literature.   
Typical of more recent work is the scheme in \cite{JLM} for  $(1+0)$-dimensional, i.e., mechanics systems in which an infinite order Lagrangian is truncated to order  $n$ as a member of a hierarchy $n \in N\to \infty$. In cases where all higher derivatives appear only in the interaction terms, this allows ``reduction" of the IVP to that of a second order system. It singles out a subset of solutions that encompass perturbation theory. This scheme was employed also in \cite{EW}
in the investigation of the problems associated with the nonlocality of string field theory vertices (cf. (\ref{deloc-b})). As pointed out in \cite{EW}, however, such schemes completely obscure the 
existence of the (infinite)  class of all other solutions exhibiting the very features one might expect associated with the nonlocal nature of the vertices such as lack of uniqueness of the IVP. 
As a general remark in this connection, finite order higher (time) derivative interactions can indeed be cast into the Hamiltonian formalism by the so-called Ostrogradsky construction. In mathematical terms this construction amounts to the usual  procedure of rewriting a $N$-th order ordinary differential equation (ODE) as a system of $N$ coupled first-order ODEs. One cannot, though, view a transcendental differential  operator (pseudo- or fractional-differential operator) as, e.g.,  (\ref{deloc-a}), as the $N=\infty$  ``limit" of this procedure. 

In this paper we consider a wide  class of nonlocal theories, which includes many of the form (\ref{deloc-a}) as a special subclass.  
Nonlocal interactions in this wide class are always defined by specifying the appropriate 
integral kernel. This includes the cases where a kernel may be associated with  a transcendental differential operator (cf. (\ref{deloc-c}) - (\ref{deloc-d}) above). Definition by the appropriate integral kernel allows a consistent, mathematically well-defined formulation in all cases. 
By the same token we eschew any series expansions in derivatives, always dealing with the complete nonlocal kernel $F(x-y)$. As implied by the preceding remarks, such expansions are generally a bad idea. They derail a 
proper posing of the classical IVP by replacing the integro-differential equations of motion by truncated higher derivative differential equations. In the quantum theory,  e.g., for various models in \cite{TT} - \cite{BK},  
unitarity can appear to be  grossly violated in inappropriate truncations at any finite order by the  
presence of increasingly large numbers of ghosts, which, however, are not there in the untruncated theory where the correct analyticity properties are regained. 
In fact, as we will see, kernels that do not even possess expansions about every point in $x$-space may comprise the physically most  promising class of nonlocal interactions.

An outline of the paper is as follows. 
In the following section \ref{Df} we introduce interactions of delocalized fields.  
We then impose conditions on the allowed interactions. Specifically, 
we give, a precise statement of the conditions imposed on the Fourier transform of allowed kernels dictated by the requirements of good UV behavior and unitarity. Among kernels that may accommodate these requirements we distinguish between strictly nonlocal and quasi-local (bounded support) kernels. 

In section \ref{IVP} we make a detailed study of the classical IVP using modern fixed-point (contraction) techniques. We first review the case of local interactions where the equations of motion are partial differential equations (PDE) giving a proof of existence and uniqueness of solutions.  We then turn to the case of nonlocal interactions specified by a given integral kernel. Varying the action now gives equations of motion that are non-linear partial integro-differential equations. This makes the contrast to the local case rather transparent. Because of the ``spilling-over" effect of the nonlocal kernel one now has an  IVP problem with both past and future ``delays". Differential and integro-differential/functional equations with past delays have been extensively studied in recent decades in a vast literature.\footnote{This is because ``real-world" modeling of many systems in physics, engineering, biology, economics and other fields very often must include dependence on past history (memory). See, e.g., \cite{B}. \label{F5}} They imply specification of functions as data over a time interval (as opposed to just data on an initial time hypersurface). In our case some specification must also be made in the future delay region. We then prove (local) existence of solutions but uniqueness is totally lost. Physics, however, is more than mathematics. If the delay regions are of strictly finite extent of length order $\ell$, they will be masked by the uncertainties in any measurements probing physics at scales larger than $\ell$. This situation arises with quasi-local interactions. 
A proper formulation of such considerations, however, can only be given in the full quantum context (section \ref{QT}). 
In a similar vein, if one assumes quasi-local interactions being asymptotically turned off (free fields in asymptotic regions) and such that a global existence result can be established, uniqueness is recovered as shown at the end of section \ref{IVP}.\footnote{Similar existence results in the case 
of the string field theory inspired kernel (\ref{deloc-a}) in $d=1$ (where it is well defined) with a ``future'' asymptotic boundary condition were obtained in \cite{PV}.  \label{Fotherex} }

In section \ref{tbH} we digress to briefly consider a  Hamiltonian formulation. The standard procedure 
gives a positive ``Hamiltonian". Due to the nonlocal interactions, however, it necessarily  contains implicit functional dependence on a range of times and thus fails to reproduce the correct equations of motion. The standard Hamiltonian procedure is simply inapplicable in the case of general nonlocal interactions. 
In the case of quasi-local (compact support) interactions, however, the results on the IVP in the previous section suggest that one should instead define a smeared Hamiltonian 
appropriate to time-blocking over intervals longer than scale $\ell$. This procedure then indeed correctly reproduces the equations of motion for evolution over such time blocks.   

In section \ref{QT} we finally turn to the quantum theory. Quantization is straightforwardly performed via the path integral. The rapid decay properties required of allowed kernels, as given in section \ref{Df},  ensure that any graph is UV finite for wide classes of scalar potentials. Similarly their analyticity properties, chosen  so as not to modify the Cutkosky cutting rules, ensure that, at least graph by graph, unitarity is preserved.  
The main concern, however, is causality.  We examine the structure of the effective propagator that results from incorporating the nonlocal kernels it joins to at vertices. This allows us to obtain a generalization of the Bogoliubov causality condition equation in the presence of nonlocal interactions. This generalized equation shows how the Bogoliubov condition for local interactions gets  modified by nonlocality. Remarkably, for interaction kernels of compact support of size $\ell$, it implies that all non causal effects remain confined within scale $\ell$. 

Some further discussion of these results and their extension and application to other theories is given in the concluding section \ref{DO}. The reader who is not interested in the details of the classical IVP can go directly from section \ref{Df} to section \ref{QT}. We work mostly in spacetime dimension $d=4$ but most considerations extend to general $d$ straightforwardly. We use standard physics and mathematical notations; in particular, $C[U]$ denotes the space of continuous functions on domain $U$, and $||\cdot ||_{L^\infty}$ denotes the $L^\infty$ norm:  $||\phi||_{L^\infty(U)} = {\esssup _{x\in  U}} |\phi(x)|$.

\section{Delocalized field interactions \label{Df}} 
\setcounter{equation}{0}
\setcounter{Roman}{0} 
We consider the simplest case of a real scalar field, the extension to complex or multi-component scalar fields being immediate. The Lagrangian with local non-derivative interactions is then given by 
\beq 
{\cal L} = {1\over 2} \partial_\mu \phi \partial^\mu \phi - {1\over 2}m^2 \phi^2 
- V(\phi)   
 \label{act1}
\eeq
with $V(\phi) \geq 0$. General polynomial interactions are allowed in $V(\phi)$, though 
it suffices keep in mind the standard $\phi^4$ example. Non-polynomial interactions, in particular interactions ensuring Lipschitz-continuity of $V$, e.g., $\phi^4 e^{-\kappa \phi^2}$, are also of interest, as we will see, in connection with the classical IVP.

The class of non-local versions of (\ref{act1}) considered in this paper is 
obtained by replacing the field $\phi$ in $V(\phi)$ by a delocalized field $\tlphi$, i.e. 
\beq 
{\cal L} = {1\over 2} \partial_\mu \phi \partial^\mu \phi - {1\over 2}m^2 \phi^2  - V(\tlphi) 
. \label{act2}
\eeq
The delocalized field $\tlphi$ is defined by 
\beq 
\tlphi(x) =\int d^4y \, F(x-y) \phi(y)   \label{deloc1}
\eeq
in terms of a delocalization kernel $F(x-y)$. The local action (\ref{act1}) is then obtained as the special case $F(x-y) = \delta^4(x-y)$.  
The kernel $F(x)$ is a scalar density function defined on the spacetime manifold $\mathbb{R}^{1+3}$, i.e., if $F$ is specified in one frame, under $x\to \Lambda x$ it is given by $F^\prime(x) = F(\Lambda^{-1} x)$. (We only need consider  $\det \Lambda=1$ here so we ignore the distinction between scalars and scalar densities.)

The kernel $F$ and its Fourier transform $\hat{F}$, introduced through 
\beq 
F(x-y) = \int d^4k \, \hat{F}(k)\,  e^{-ik\cdot(x-y)} \, ,  \label{deloc2} 
\eeq
will be the quantities of main concern below.

It should be noted that in any interaction term in $V(\tlphi)$, such as $ \tlphi^n(x)/n!$,  the delocalized fields $\tlphi$ interact at a common spacetime point $x$, resulting in an interaction  $\int \prod_{i=1}^n d^4y_iF(x-y_i) \phi(y_i)$. This ensures that each interaction vertex is proportional to a single spatial momentum conservation delta-function (times an energy 
conservation delta-function) which, as it is well-known, is a sufficient condition for cluster decomposition.  This would not necessarily be the case for generic nonlocal interactions  
\beq 
{\cal L}_n(x) ={\lambda_n\over n!}\, \int \prod_{i=1}^n d^4x_i\,F(x, x_1, \ldots, x_n)\,\phi(x_i)  \, .  \label{act3}
\eeq
Feynman rules for diagrammatic expansion can, of course, be read off the non-local Lagrangians (\ref{act2}) - (\ref{deloc1}), or 
(\ref{act3}), \cite{tHV1} just as in the local case (\ref{act1}).

In what follows we adopt a general point of view and consider a wide class of 
delocalization integral kernels. 
Kernels in this wide class cannot necessarily be represented by an expansion in derivatives about every point, those associated with operators of the form (\ref{deloc-a}) being a particular subclass of possible non-local interactions. 
In general, the choice of the kernel $F(x)$ in (\ref{deloc1}) is restricted by the imposition of 
physical requirements. The first two requirements are controllable UV behavior and unitarity, at least within the perturbative expansion. These conditions constrain $\hat{F}(k)$. 
Causality and other considerations may then further guide the choice of $F$ in particular theories. 
 
The two fundamental properties we require $\hat{F}(k)$ to satisfy are:
\begin{enumerate}
\item[(I)] $\hat{F}(k^)$ is an element of the space of {\it functions of rapid decay}, commonly denoted by $C^\infty_\downarrow(\mathbb{R}^{1+3})$. 

A function $\psi(u) \in C^\infty_\downarrow(\mathbb{R}^{1+3})$ is: (i) infinitely differentiable; (ii) 
\beq 
\sup_u \,\left| u^p D^q \psi(u) \right| \,   < \,  \infty   \label{rdf1}
\eeq
for every pair of nonnegative integer multi-indices $p,q$, with the notation  $u^p \equiv  u_0^{p_0} \cdots u_3^{p_3}$ and $D^q \equiv  \partial^{q_0 + \cdots + q_3} / \partial u_0^{q_1}\cdots \partial u_3^{q_3}$. Note that the boundedness of $u^{p+1} D^q\psi(u)$ shows that 
\beq 
\lim_{||u||\to \infty} |u^p D^q\psi (u)| \to 0   \, ,           \label{rdf2}
\eeq 
i.e.,  a function of rapid decay and all its derivatives in fact vanish faster than any negative power of its arguments as $|u|\to \infty$.

\item[(II)] $\hat{F}(k)$ is an entire analytic function, i.e. it is analytic in each component $k_\mu$ in the entire complex $k_\mu$-plane. 
\end{enumerate}
Condition (I) is dictated by the requirement of good UV behavior, and condition (II) by that of unitarity. 

We now define two  classes of vertices constructed from products of delocalized fields (\ref{deloc1})
with kernels that can accommodate the conditions (I), (II) above:
\begin{enumerate}
\item[(a)]
{\it Quasi-local interaction vertices} with integral kernels $F$ which are elements of the space of {\it functions of bounded support}, commonly denoted by $C^\infty_c(\mathbb{R}^{1+3})$.

A function $\psi\in C^\infty_c(\mathbb{R}^{1+3})$: (i) is infinitely differentiable; (ii) vanishes identically outside a compact (bounded, closed) set, i.e. has bounded support (the closure of the set on which the function has non-zero value).  
\item[(b)] {\it Strictly nonlocal interaction vertices} 
with integral kernels $F$ which are elements of $ C^\infty_\downarrow(\mathbb{R}^{1+3})$ but not of $C^\infty_c(\mathbb{R}^{1+3})$, i.e. elements of the space of functions of rapid decay that do not have compact support. 
\end{enumerate} 
An example of a nonlocal kernel, characterized by a length scale $\ell$,  would be $F(x) = \exp {- [x^2/\ell^2]^2 }$. 
Quasi-local kernels are naturally obtained by standard smoothing (`mollifying') of $\delta^4(x)$ over a length scale $\ell$.  
In other words, they may be taken as elements of a $C^\infty_c$ delta-family, the local  limit being recovered when $\ell\to 0$. 
It is important to note here that a quasi-local kernel of such or similar shape of size $\ell$ will have a Fourier transform which is essentially flat for real momenta less than of order $1/\ell$, and rapidly decaying for momenta above $1/\ell$; in the local limit  $\ell\to 0$ one recovers $\hat{F}(k) =$ constant 
for all $k$. A nonlocal kernel will have a Fourier transform of approximately similar behavior if the kernel decays sufficiently fast outside a sufficiently small region of size $\ell$. It should also be noted that there are points about which quasi-local kernels, though $C^\infty$ functions, have Taylor expansions of zero radius of convergence.

We now recall some mathematical facts (see, e.g., \cite{GS}).
$C^\infty_c$ supplemented with an appropriate notion of convergence becomes ${\mathscr D}(R)$, the test function space of distributions, the latter being defined as elements of  ${\mathscr D}^\prime(R)$, the dual space to ${\mathscr D}(R)$. $C^\infty_\downarrow$ supplemented with an appropriate notion of convergence becomes ${\mathscr S}(R)$, the test function space of tempered distributions which are elements of the dual space ${\mathscr S}^\prime(R)$.  
Thus our delocalized fields replace local fields (tempered distributions in the usual formalism) by their convolution with an element of the corresponding test function (sub)space.

A basic result for us here is that if a function is in $C^\infty_\downarrow$ its Fourier transform exists and is also in $C^\infty_\downarrow$.  Now, $C^\infty_c$ is  a subspace of the space of functions of rapid decay $C^\infty_\downarrow$. Note, however, that the Fourier transform of a function in $C^\infty_c$ will not be also in $C^\infty_c$, though, of course, it will in $C^\infty_\downarrow$. 
(This is in fact what necessitates the introduction of $C^\infty_\downarrow$ in order to have the distributional extension of classical Fourier theory.)
It follows that the Fourier transforms of kernels of both types (a) and (b) above satisfy condition (I).

Another basic fact is that the Fourier transform of a function in $C^\infty_c$ is an entire analytic function of its argument. Thus, the Fourier transforms of functions in $C^\infty_c$ form a subset of $C^\infty_\downarrow$ consisting of functions that can be extended from the real axis to the complex domain as  entire functions.\footnote{Given appropriate norms, these functions comprise the sequence of $Z$ spaces of entire functions in the Gel'fand-Shilov extension of generalized function theory into the complex plane \cite{GS}. They possess special properties of potential interest in particular models of interactions, but we will not make specific use of them in the general considerations in this paper. \label{FZspace} } It follows that the Fourier transforms $\hat{F}(k)$ 
of quasi-local kernels (a) above satisfy also condition (II). In contrast, condition (II) is not automatically satisfied for general non-local kernels (b), i.e. general elements of $C^\infty_\downarrow$, though it can be satisfied for large subclasses of them, (\ref{deloc-d}) being a simple example.

Within this framework, operators of the form (\ref{deloc-a}) are considered as defined in terms of the corresponding integral kernels, as exemplified by the r.h.s. in (\ref{deloc-b}), (\ref{deloc-c}). 
Thus, (\ref{deloc-d}), being indeed of rapid decay, gives a non-local interaction of type (b) above. 
Note also that it may not always be possible to give a closed kernel form in both coordinate and $k$-space, but being in $C^\infty_\downarrow$ in either space is sufficient.   
In contrast, in the case of (\ref{deloc-b}) for $d\geq 2$ the integral kernel 
is not of rapid decay for large negative values of its argument $(x-y)^2$, and is not included in the class (b); its Fourier transform cannot be properly defined to satisfy (II) with the consequent lack of analyticity presumably implying instability.  

In general, assuming they are well-defined by appropriate choice of the function $f$, 
from among the kernels that may be introduced via (\ref{deloc-a}) a large subset are of the non-local type (b). Another subset though, e.g., some of those employed in \cite{TT}, also \cite{Mod1}, though well-defined, are not of rapid decay and do not belong to the class (b). On the other hand, general kernels of type (b) constitute a rather larger set than the rapid decay subset that can be generated via (\ref{deloc-a}). Quasi-local vertices of type (a) cannot, of course, be generated via (\ref{deloc-a}).

In summary, we introduced non-locality by the replacement of fields in local interaction vertices by delocalized fields, i.e., replacement of (\ref{act1}) by (\ref{act2}) - (\ref{deloc1}). The Fourier transform of the delocalization kernel is required to satisfy the fundamental conditions (I) and (II) above. We then defined two general classes of kernels, the non-local kernels of rapid decay, and the quasi-local kernels of bounded support. Both satisfy condition (I). Quasi-local kernels also satisfy 
condition (II). General non-local kernels do not automatically satisfy condition (II), but a large subset in this class does. In what follows we examine the extent to which these classes of interactions may allow physically viable theories.

\section{The classical theory - time evolution and IVP \label{IVP}}
\setcounter{equation}{0}
\setcounter{Roman}{0} 

The equation of motion is derived as usual by varying the action (\ref{act2}). 
The resulting field equation is 
\beq 
\Box \phi(x)  + m^2\phi(x) + \int d^d z\, F(x-z) V^{\, \prime}(\tlphi(z))   = 0 \, , \label{eom1}
\eeq
with, as usual, $V^\prime(w) \equiv {d V(w) / d w}$. 
(\ref{eom1}) is a nonlinear wave functional equation in $d$ spacetime dimensions. The functional field dependence introduces  ``delays".  
Differential/functional equations with past delays have been extensively studied in recent years. The novel feature here is that the delocalized field interactions introduces future, as well as past, delays. We need then examine what effect this has on time evolution and the existence of solutions 
of (\ref{eom1}).  In this section we set $m=0$ since this simplifies formulas without affecting 
the main issues we discuss here - the non-zero mass case is quite analogous.\footnote{In any case, one may always include the mass term as a local addition to the potential as is usually done in the mathematical literature.\label{Fmass}}

A common approach to existence and uniqueness questions for nonlinear wave equations is to convert them to an integral equation. This is done by 
substituting the non-linear interaction for the inhomogeneous term in the known solutions 
to the linear inhomogeneous problem.  In the case of (\ref{eom1}) 
with $m=0$ in $d=2$ ($x =(x^1,t)$) d'Alembert's formula results in the integral equation 
\bea
\phi(x^1, t) &=& \varphi(x^1,t) - {1\over 2} \int_0^t ds \int_{x^1-t+s}^{x^1+t-s} dy \int d^2z \;F(y - z^1, s- z^0))  
V^{\, \prime} (\tlphi(z^1,z^0))  \nonumber  \\
& = & \varphi(x^1,t) - {1\over 2} \int_{D(x^1,t)} ds \, dy  \int d^2z \; F(y - z^1, s- z^0))  \,
\; V^{\, \prime}(\tlphi(z^1,z^0))  \,  ,\label{ie1}
\eea 
whereas in $d=4$ ($x=(\bfx,x^0)$) Kirchhoff's formula leads to  
\beq 
\phi(\bfx,t)  =  \varphi(\bfx,t) - {1\over 4\pi} \int_{B(\bfx,t)} d^3 \bfy \int d^4z 
 \; {F(\bfy - \bfz, t - |\bfx - \bfy| - z^0) \over |\bfx - \bfy|} \;
 V^{\, \prime}(\tlphi(\bfz,z^0))  \, . \label{ie2}
\eeq
In (\ref{ie1}) and (\ref{ie2}) $\varphi$ denotes the solution, in $d=2$ and $4$, respectively, to the linear homogeneous problem 
\beq
\Box \varphi(\bfx,t) =0  \, . \label{eom2}
\eeq
In (\ref{ie1}) $D(x^1,t)$ denotes the triangular region (domain of dependence) enclosed by the $t=0$ axis and the two characteristics lines $x^1\pm t$ emanating from the point $(x^1,t)$. In (\ref{ie2}) $B(\bfx,t)= \{ \bfy \in \mathbb{R}^3 \mid |\bfx-\bfy| \leq t \}= \{$ the closed ball in $\mathbb{R}^3$ centered at $\bfx$ and of radius $t > 0\}$.\footnote{Note the amusing fact that whereas $D(x^1,t)$ is $2$-dimensional, $B(\bfx,t)$ is only $3$-dimensional. This is because Huygen's principle holds in  odd space dimensions $(d-1)\geq 3$ but not for $(d-1)=1$ - the latter behaving like even space dimensions. \label{FHuyg}} The presence of the unrestricted $z$-integration in (\ref{ie1}) and (\ref{ie2}) is due to the nonlocal nature of the kernel $F$, and is the crucial feature distinguishing the nonlocal from the local case.

The integral equation formulation allows application of fixed point theorems to investigate existence of solutions \cite{SS}. In an appropriately defined space of functions $X$ (cf. below) 
the r.h.s. of (\ref{ie1}) or  (\ref{ie2}) defines a nonlinear mapping $A: X\to X$.  
A fixed point of this mapping, i.e. 
\beq 
\phi(\bfx,t) = A[\phi](\bfx,t)   \, ,        \label{ie3}
\eeq 
if it exists, is then a solution to these integral equations. This formulation also 
makes immediately apparent the difference between the local and nonlocal cases. 

\subsection{Local interaction \label{IVP-l}} 
It will be useful to first review here an existence and uniqueness proof of this kind \cite{SS} in the familiar case of local interactions, i.e. (\ref{ie2}) with $F(x-y)=\delta^4(x-y)$.  The r.h.s. of (\ref{ie2}) then defines the mapping:  
\beq 
A[\phi](\bfx,t)  =  \varphi(\bfx,t) - {1\over 4\pi} \int_{B(\bfx,t)} d^3 \bfy 
 \; {1 \over |\bfx - \bf y|} \;
V^{\,\prime}(\phi(\bfy, t_r))  \, 
 \label{A1}
\eeq
with $ t_r = t - |\bfx - \bf y|$. The solution $\varphi$ to the linear problem (\ref{eom2}) 
satisfies initial data 
\beq 
\varphi(\bfx,0) = g(\bfx) \, , \qquad  \partial_t \varphi (\bfx,0) = h(\bfx) \, . \label{eom3}
\eeq
We then seek solutions to (\ref{ie3}) for (\ref{A1}). 

\begin{figure}[ht]
\begin{center}
\includegraphics[width=0.7\textwidth]{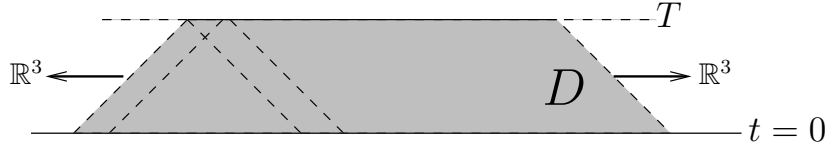}
\end{center}
\caption{Local nonlinear wave equation IVP on domain 
$D \to [0,T]\times \mathbb{R}^3$ stitched together from the conical dependence domains of (\ref{A1}) for points on $t=T$ hypersurface.      \label{qlF1}}
\end{figure}

Let $X$ denote the set of functions 
\[ X \equiv \big\{ \phi \in C([0,T]\times \mathbb{R}^3) \; \; \big| \;\; \phi(\bfx,0)= g(\bfx), \, ||\phi - \varphi||_{L^\infty} \leq 1 \big\}\, , \label{Xa} \]
a subset of the complete metric space ${\cal S} \equiv \{ \phi \in C([0,T]\times R^3) 
\mid 
\phi(\bfx,0)= g(\bfx) \}$. Assuming smooth $g, h$ in (\ref{eom3}), $\varphi$ is smooth, and it follows that if $\phi \in X$ there exists a constant $C_0$ such that 
\beq 
||\phi||_{L^\infty([0,T]\times\mathbb{R}^3)} \leq C_0  \, .   \label{X1}
\eeq
For the mapping (\ref{A1}) then, if $\phi, \psi \in X$ one has 
\bea 
|| A[\phi] - A[\psi] ||_{L^\infty([0,T]\times\mathbb{R}^3)}  
      &  \leq &   \sup_{x\in R^3, 0\leq t\leq T} \left({1\over 4\pi} \int_{B(\bfx, t)} 
  \!  \!\!  d^3 \bfy 
 \; {1 \over |\bfx - \bf y|} \; \left|
V^{\,\prime}(\phi(\bfy, t_r))  - V^{\,\prime}(\psi(\bfy, t_r))  \right| \right) \,   \nonumber \\
& \leq &    C_1   \sup_{x\in R^3, 0\leq t\leq T} \left({1\over 4\pi} \int_{B(\bfx, t)} 
  \!  \!\!  d^3 \bfy 
 \; {1 \over |\bfx - \bf y|} \; \left| \phi(\bfy, t_r) - \psi(\bfy, t_r) \right| \right)  \nonumber \\
& \leq & C_1 || \phi - \psi||_{L^\infty([0,T]\times\mathbb{R}^3)}  \sup_{x\in R^3, 0\leq t\leq T} \left({1\over 4\pi} \int_{B(\bfx, t)} 
  \!  \!\!  d^3 \bfy  \; {1 \over |\bfx - \bf y|}\right)   \nonumber \\
  & \leq & {1\over 2} C_1 \,T^2 \, || \phi - \psi||_{L^\infty([0,T]\times\mathbb{R}^3)}  \, . \label{A1ineq1} 
\eea
The constant $C_1$ comes from 
\bea
\left|V^{\,\prime}(\phi(\bfy, t_r))  - V^{\,\prime}(\psi(\bfy, t_r))  \right| &  \leq & 
\max_{|w|\leq C_0} |V^{\,\prime\prime}(w)| \,  \left| \phi(\bfy, t_r) - \psi(\bfy, t_r) \right|  \nonumber \\
& \leq &  C_1 \,  \left| \phi(\bfy, t_r) - \psi(\bfy, t_r) \right|  \, , \label{A1ineq2}
\eea 
where it is assumed that $V(w)$ is at least twice continuously differentiable and use is made of (\ref{X1}), and also 
$\left({1\over 4\pi} \int_{B(0, R)}  \!  d^3 \bfy  \; {1 \over |\bf y|}\right) =  R^2/2$.    
Take  $T$ small enough and/or couplings in $V$ weak enough so that $ {1\over 2}C_1T^2 <1$. It follows then from (\ref{A1ineq1}) that the mapping $A$ is a contraction. Similarly, if $\phi \in X$, 
\bea 
|| A[\phi] - \varphi ||_{L^\infty([0,T]\times\mathbb{R}^3)}  
      &  \leq &   \sup_{x\in \mathbb{R}^3, 0\leq t\leq T} \left({1\over 4\pi} \int_{B(\bfx, t)} 
  \!  \!\!  d^3 \bfy 
 \; {1 \over |\bfx - \bf y|} \; \left|
V^{\,\prime}(\phi(\bfy, t_r)) \right| \right) \,   \nonumber \\
 & \leq & {1\over 2} C_1  \,T^2 \, || \phi||_{L^\infty([0,T]\times\mathbb{R}^3)}  \nonumber \\
& \leq & {1\over 2} C_0 C_1  \,T^2  \, . \label{A1ineq3} 
\eea 
so that indeed, under the same conditions, $A: X \to X$.

Starting then with some initial field configuration $\phi_0\in X$, 
one obtains a succession of configurations by 
\beq 
\phi_{n}  = A[\phi_{n-1}]  \, . \label{ie4}
\eeq  
By the contraction mapping theorem this has a unique fixed point  which provides the unique solution to (\ref{ie3}) - (\ref{A1}) on $D= [0,T]\times \mathbb{R}^3$ (Fig \ref{qlF1}).  Further arguments \cite{SS}, in particular consideration of the corresponding integral equations for derivatives, show that, assuming $V$ is smooth,  all 
derivatives are similarly bounded, so the solution is in fact smooth. Depending on the form of the interaction $V$ continuation of the solution to any $T$, i.e. global existence,  may be proven in many cases (such as a $\phi^4$ interaction). 
In fact, under the stronger assumption that the interaction is Lipschitz continuous it may be proven that a weak solution always exists for any $T$.

\subsection{Nonlocal interaction \label{IVP-nl}} 
\begin{figure}[ht]
\begin{center}
\includegraphics[width=0.7\textwidth]{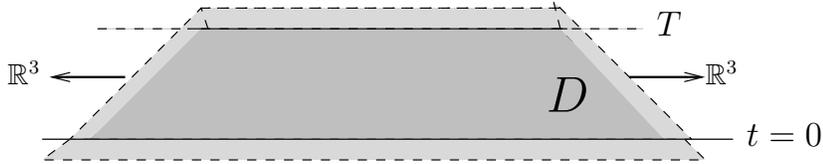}
\end{center}
\caption{Nonlocal version of the IVP on domain $D$, now bordered by past and future delay regions (light shading) due to ``spilling over" by the nonlocality of the interaction. The delays are of finite width for kernels of bounded support as shown here. \label{qlF2}}
\end{figure}

In the nonlocal case $A[\phi]$ is given by the r.h.s. of (\ref{ie2}) involving nonlocal kernel $F$:
\beq 
A[\phi] =    \varphi(\bfx,t) - {1\over 4\pi} \int_{B(\bfx,t)} d^3 \bfy \int d^4z 
 \; {F(\bfy - \bfz, t - |\bfx - \bfy| - z^0) \over |\bfx - \bfy|} \;
 V^{\, \prime}(\tlphi(\bfz,z^0))  \, . \label{A2}
\eeq
In considering (\ref{ie4}) the difference from the local case is now apparent. It is no longer true that 
to compute $\phi_{n+1}$ one needs $\phi_n$ only in $D$. 
There is a ``spill-over" or ``delay" effect 
since in (\ref{ie2}) $(\bfx,t)$ is connected to  $(\bfz,z^0)$ which now can range outside $B(\bfx,t)$. 
Similarly, in the $d=2$ case, on the r.h.s. in (\ref{ie1})  the point $(x^1,t)$ is connected via the kernel 
to $(z^1,z^0)$ which can range outside $D(x^1,t)$. 
Note that there is a past as well as a future delay. The delay regions then form a collar in $\mathbb{R}\times \mathbb{R}^3\setminus D$ (Fig. \ref{qlF2}). For kernels of bounded support this collar  is of finite extent. For nonlocal kernels this extent is strictly infinite even though, for sufficiently rapid decay rate, it may appear essentially finite for all practical purposes.

\subsubsection{Quasi-local interactions}
We now restrict to the case of bounded support kernels of scale $\ell$. 
Let $\tilde{D}^{(k)}= D\cup D_c^{(k)} $ be the extension of the domain $D$ to include a collar region $D_c^{(k)}$
of width $2lk$ with integer $k\geq 1$.  
Given some initial configuration $\phi_0$, to compute 
$\phi_1$ in $D$ via the mapping $A$ one needs $\phi_0$ in $\tlD^{(1)}$; to compute 
$\phi_2$ in $D$ one needs $\phi_1$ in $\tlD^{(1)}$, hence $\phi_0$ in $\tlD^{(2)}$, and so on: 
for $n > m \geq 0$, to compute $\phi_n$ in $D$ one needs $\phi_m$ in $\tlD^{(n-m)}$. 
Since, to apply a fixed point argument, $n \to \infty$ one sees that the field need be computed at essentially all points. 
This may at first sight appear not to be a problem since, given some initial configuration $\phi_0$, say,  identically vanishing except inside some  bounded spacetime region, one may apply the mapping $A$ to generate  the field value of subsequent configurations at any spacetime point. Because $A$ involves integration, however, it is easily seen that upon subsequent iterations it will generate fields that grow as $t^2$ at spacetime point $x=(\bfx,t)$. This was already seen in the local interaction case above; there, however, for solutions on a fixed interval $0\leq t \leq T$ only the values of fields within this interval were needed in successive iterations. In contrast, in the case of nonlocal interaction to establish existence of solutions on $[0, T]$ in the same manner would require field values at all points with the attendant boundedness problem. We must then proceed differently. 


The existence of local solutions on suitably small intervals can, in fact, be demonstrated 
in the case of nonlocal interaction, but uniqueness is completely lost. To demonstrate existence on $[0,T] \times \mathbb{R}^3$ one has to extend the definition of the mapping $A$ to specify fields in the collar region  $D^{(1)}_c\equiv D_c$. 
In the past delay region of the collar, $\{x\in D_c \cap \{t\leq 0\}\}$, the fields  
are specified as initial data (cf. \cite{B}). One may take\footnote{Alternatively, one may specify an arbitrary smooth function $\psi$ on $-t_0 \leq t \leq 0$ as past delay initial data, and replace (\ref{eom4}) with 
$\varphi (\bfx, 0) = \psi(\bfx, 0)$ and  $\partial_t \varphi (\bfx, 0) = \partial_t \psi(\bfx,0)$.\label{F6}} 
the solution of the linear problem (\ref{eom2}) 
with initial data 
\beq 
\varphi (\bfx, -t_0) = g(\bfx) \, , \qquad  \partial_t \varphi (\bfx, -t_0) = h(\bfx) \, . \label{eom4}
\eeq
at some time $t=-t_0$ where $t_0 \geq 2l$, and set $\phi(x) = \varphi(x)$ on $-t_0 \leq t \leq 0$. 
Let 
\beq 
\tlA[\phi] (x) = \left\{ \begin{array}{l l l} 
A_c[\phi](x) & \mbox{for} \quad x\in D_c \cap \{t >  0\} \\ 
A[\phi](x) & \mbox{for} \quad x\in D \\
\varphi(x)  & \mbox{for} \quad  x\in  D_c \cap \{t <  0\}  \end{array} \right.  \, .   \label{A3}
\eeq
$A_c[\phi]$ is specified by extending the values $\phi(x),  x\in \partial D\cap \{t >  0\}$, 
given by $A[\phi]$ into the collar region $D_c \cap \{t >  0\}$ in some prescribed continuous fashion. 
E.g., parametrizing points in  $D_c \cap \{t >  0\}$  by 
\beq 
x= y+ s \nu(y), \quad s \in [0,2l],    \quad y\in \partial D  \, ,  \label{A4}
\eeq 
where $\nu^\mu(y)$ denotes the outward (timelike) normal at $y\in \partial D$, 
we may define 
\beq 
A_c[\phi](x) = \zeta(x) [ A[\phi](y) - \varphi(y)]  + \varphi(x)  \, . \label{A5}
\eeq
In (\ref{A5}) $\zeta(x)$ is a $C^\infty_c([0,2l]\times \mathbb{R}^3)$ function such that $0\leq \zeta \leq 1$ and $\zeta(x) \equiv 1$ in a neighborhood of $\partial D\cap \{t >  0\}$. 
Adopting one such prescription, (\ref{A3}) defines then a mapping $\tlA$ on $\tlD = D\cup D_c$.

Let $X$ denote the set of functions 
\[ X \equiv \big\{ \phi \in C([-2l,T+2l]\times \mathbb{R}^3) \;  \big| \; \phi(x)= \varphi(x)\  \mbox{on} \ 
-t_0 \leq t \leq 0, \;||\phi - \varphi||_{L^\infty([0,T+2l] \times \mathbb{R}^3)} \leq 1 \big\}\, . \label{Xb} \]
Assuming smooth $g, h$ in (\ref{eom4}), it follows that if $\phi \in X$ there exists a constant $C_0$ such that 
\beq 
||\phi||_{L^\infty([0,T+2l] \times \mathbb{R}^3)} \leq C_0  \, .  \label{X2}
\eeq
Furthermore, 
\beq 
||\phi||_{L^\infty([a,b]\times \mathbb{R}^3)} = \sup_{a\leq t \leq b, \ \bfx\in \mathbb{R}^3} \left| \int d^4 y  
F(x-y)\phi(y) \right| \leq C\, || \phi ||_{L^\infty ( [a-l, b+l]\times \mathbb{R}^3)} \label{X3} 
\eeq
where   $C= \int d^4u |F(u) |$. 

If $\phi, \psi \in X$ one has 
\bea 
\sup_{{\s 0 \leq z^0  \leq T+l} \atop {\s \bfz\in \mathbb{R}^3}} \left| V^{\,\prime}(\tlphi(\bfz, z^0))  - V^{\,\prime}(\tlpsi(\bfz, z^0))  \right|  
&\leq & \max_{|w|\leq C_0} |V^{\,\prime\prime}(w)| \,  
\sup_{{\s  0 \leq z^0  \leq T+l} \atop {\s \bfz\in \mathbb{R}^3}}
\left| \tlphi(\bfz, z^0) - \tlpsi(\bfz, z^0) \right|   \nonumber \\
&\leq & C_1 C \, || \phi - \psi ||_{L^\infty([-2l, T+2l]\times \mathbb{R}^3) } \, .  \label{A2ineq1} 
\eea
Note also that $|| \phi - \psi ||_{L^\infty([-2l, T+2l]\times \mathbb{R}^3) }= || \phi - \psi ||_{L^\infty([0, T+2l]\times \mathbb{R}^3) }$.   
Making then use of (\ref{X2}), (\ref{X3}) and (\ref{A2ineq1}), 
\bea 
|| A[\phi] - A[\psi] ||_{L^\infty([0,T] \times \mathbb{R}^3)}  
      &  \leq &   \sup_{0\leq t\leq T, \ \bfx\in \mathbb{R}^3} \Bigg({1\over 4\pi} \int_{B(\bfx, t)} 
  \!  \!\!  d^3 \bfy \int d^4z
 \; {F(\bfy - \bfz, t_r - z^0)  \over |\bfx - \bfy|}   \nonumber \\
  & & \qquad  \hspace{2.5cm} \cdot  \left|
V^{\,\prime}(\tlphi(\bfz, z^0))  - V^{\,\prime}(\tlpsi(\bfz, z^0))  \right| \Bigg) \,   \nonumber \\
& \leq &     \sup_{0\leq t\leq T, \ \bfx\in \mathbb{R}^3} {1\over 4\pi} \int_{B(\bfx, t)} 
  \!  \!\!  d^3 \bfy \int d^4 z 
 \; {F(\bfy - \bfz, t_r - z^0)  \over |\bfx - \bfy|} \;  \nonumber \\
 & & \qquad \hspace{2.5cm} \cdot \,  C_1 C \, || \phi - \psi ||_{L^\infty([-2l, T+2l]\times \mathbb{R}^3) }
\nonumber \\
& \leq & {1\over 2} C_1 C^2 \, T^2 \, || \phi - \psi ||_{L^\infty([-2l, T+2l]\times \mathbb{R}^3) } 
\, .  \label{A2ineq2}
\eea
Also, for $x\in D_c \cap \{t> 0\}$, 
\beq 
\Big| A_c[\phi](x) - A_c[\psi](x) \Big| \leq \Big| A[\phi](y) - A[\psi](y) \Big|  \, .  \label{A2ineq3}
\eeq
Thus, for $\phi, \psi$ in $X$, 
\beq 
|| \tlA[\phi] - \tlA[\psi] ||_{L^\infty([0,T+2l] \times \mathbb{R}^3)}  
 \leq  {1\over 2} C_1 C^2 \, T^2 \, || \phi - \psi ||_{L^\infty([0, T+2l]\times \mathbb{R}^3) }  \, . \label{A2ineq4} 
\eeq 
Similarly, one verifies that 
\beq 
|| \tlA[\phi] - \varphi ||_{L^\infty([0,T+2l] \times \mathbb{R}^3)}  
 \leq  {1\over 2} C_0 C_1 C^2 \, T^2 \, . 
 \label{A2ineq5}
 \eeq
Hence, for $T$ and/or couplings in the interaction $V$ chosen sufficiently small, $\tlA: X\to X$ and is a contraction.  It follows that there is a solution to the integral equation (\ref{ie2})  on $[0,T]$ satisfying the initial data on $ [-t_0, 0]$. 

The solution is unique for a given mapping $\tlA$, but this mapping   
depends on its specification in the future collar region $D_c$.  There is then a unique solution on $[0,T]$ for each such choice of specification. In other words, {\it there is local existence but complete loss of uniqueness} for the solution of the  IVP associated with (\ref{ie2}) (or (\ref{ie1})).  
{\it This reflects the loss of causality due to the nonlocality of the interaction.} To obtain a solution on $[0,T]$, in addition to initial data on some past delay region $[-t_0,0]$,  some specification of fields in the future delay region $[T, T+ 2l]$ has to be provided, thus converting the IVP to a combined IVP-BVP of sorts.

For physics at length scales appreciably longer than $\ell$, however, this dependence on the acausal delay from $[T, T+ 2l]$ would be masked 
in field measurements necessarily averaged over scales longer than order $\ell$. This becomes even more pertinent  when going from the classical to the quantum context where all fluctuations
are included (section \ref{QT}). 
This suggests that one should consider time evolution smeared over regions of order $l$. We return to this below.

There are circumstances in which  
uniqueness of the classical IVP can be obtained. This is the case when the nonlocal interactions are confined  within the interval $[0,T]$ for which existence can be demonstrated. This is of particular importance if existence  for any $T$, i.e. global existence (at least of a weak solution) can be proven. One may then consider the asymptotic ``switching-off of interactions" outside a (large) spacetime region, as commonly done in scattering theory. This is  implemented by 
letting $V(\phi)  \to g(x)V(\phi)$ where $g(x)$ is a $C^\infty$ function such that  $g(x)\equiv 1$ for $|x^0|\leq T_0-\epsilon$ and $g(x) \equiv 0$ for $|x^0 | \geq  T_0+\epsilon$.  In the rest of this section we redefine $V$ to include the variable coupling $g(x)$.  

To obtain a global existence result we will assume that $V^\prime$ is Lipschitz-continuous, i.e., 
\beq 
| V^\prime(\phi) - V^\prime(\psi) | \leq  C_1 |\phi - \psi|  \, , \label{Lcon} 
\eeq 
for some constant $C_1$. 
We also introduce the norm (cf., e.g., \cite{B}, \cite{D}) 
\beq 
|| f ||_{L^\infty_\rho([a,b] \times \mathbb{R}^3)}   = \esssup_{a\leq t\leq b, \ \bfx\in \mathbb{R}^3}   e^{-\rho t} |f(t, \bfx) |  \, .   \label{rhonorm}
\eeq
Let $X$ denote the set of functions 
\[ X \equiv \big\{ \phi \in C([-t_0,T]\times \mathbb{R}^3) \;  \big| \; \phi(x)= \varphi(x)\  \mbox{on} \ 
-t_0 \leq t \leq 0\, , \quad T> T_0+\epsilon 
\big\}\, .   \label{Xc} \]
If $\phi, \psi \in X$,  one now has 
\bea 
\left| V^{\,\prime}(\tlphi(\bfz, z^0))  - V^{\,\prime}(\tlpsi(\bfz, z^0))  \right|  
&\leq & C_1 \,  
\left| \tlphi(\bfz, z^0) - \tlpsi(\bfz, z^0) \right|   \nonumber \\
&\leq & C_1  \int dw^4 \,\left|F(\bfz - \bfw, z^0 - w^0)\right|  \left| \phi(\bfw, w^0) - \psi(\bfw, w^0) \right|   \nonumber \\
& \leq & C_1\,  e^{\rho z^0}\int  dw^4 \, 
\left|F(\bfz - \bfw, z^0 - w^0)\right| e^{-\rho (z^0 - w^0)}  \nonumber \\
& & \hspace{1.5cm}  \cdot \,  || \phi - \psi ||_{L^\infty_\rho([0,T] \times \mathbb{R}^3)}   
\nonumber \\
&\leq & C_1 C  e^{\rho z^0} e^{\rho l} \,  || \phi - \psi ||_{L^\infty_\rho([0,T] \times \mathbb{R}^3)}  
\, ,  \label{A2ineq6} 
\eea
using (\ref{Lcon}) and the fact that, in the next to last inequality, $|z^0- w^0| \leq l$ due to the bounded support of $F$. 
For  (\ref{ie2}), with initial data (\ref{eom4}),  we define the mapping $\tlA$ by 
\beq 
\tlA[\phi] (x) = \left\{ \begin{array}{l l l} 
A[\phi](x) & \mbox{for} \quad x\in D \\
\varphi(x)  & \mbox{for} \quad  -t_0 \leq t \leq 0   \end{array} \right.  \, .   \label{A6}
\eeq
Then 
\bea 
|| A[\phi] - A[\psi] ||_{L^\infty_\rho([0,T] \times \mathbb{R}^3)}  
      &  \leq &   \sup_{0\leq t\leq T, \ \bfx\in \mathbb{R}^3} \Bigg({1\over 4\pi}  e^{-\rho t} \int_{B(\bfx, t)} 
  \!  \!\!  d^3 \bfy \int d^4z
 \; {\left|F(\bfy - \bfz, t_r - z^0)\right|  \over |\bfx - \bfy|}   \nonumber \\
  & & \qquad  \hspace{2.5cm} \cdot  \left|
V^{\,\prime}(\tlphi(\bfz, z^0))  - V^{\,\prime}(\tlpsi(\bfz, z^0))  \right| \Bigg) \,   \nonumber \\
& \leq &     \sup_{0\leq t\leq T, \ \bfx\in \mathbb{R}^3} {1\over 4\pi} C_1 C \, e^{\rho l}   e^{-\rho t}  \int_{B(\bfx, t)} 
  \!  \!\!  d^3 \bfy \int d^4 z 
 \; {\left|F(\bfy - \bfz, t_r - z^0)\right|  \over |\bfx - \bfy|} \;  \nonumber \\
&  &  \hspace{2.5cm} \cdot  \,  e^{\rho z^0} \,  || \phi - \psi ||_{L^\infty_\rho([0,T] \times \mathbb{R}^3)}  
\nonumber \\ 
 &  \leq &      \sup_{0\leq t\leq T, \ \bfx\in \mathbb{R}^3} {1\over 4\pi} C_1 C  e^{\rho l}   e^{-\rho t}  \int_{B(\bfx, t)} 
  \!  \!\!  d^3 \bfy \int d^4 z 
 \; { \left| F(\bfy - \bfz, t_r - z^0) \right|  \over |\bfx - \bfy|} \nonumber \\
 & & \hspace{3cm} \cdot \, e^{\rho t_r} e^{\rho l} \,  || \phi - \psi ||_{L^\infty_\rho([0,T] \times \mathbb{R}^3)}    \nonumber 
 \eea 
using (\ref{A2ineq6}) and, again, that $|t_r - z^0|\leq l$ due to the compact support of $F$. Since $t_r = t - |\bfx - \bfy|$ we finally obtain 
 \bea 
 || A[\phi] - A[\psi] ||_{L^\infty_\rho([0,T] \times \mathbb{R}^3)} 
 & \leq &  \sup_{0\leq t\leq T, \ \bfx\in \mathbb{R}^3} {1\over 4\pi} C_1 C^2\,  e^{2\rho l}    \int_{B(\bfx, t)} 
  \!  \!\!  d^3 \bfy  \;  { e^{- \rho |\bfx-\bfy| } \over |
  \bfx - \bfy|  }\, 
  \,  || \phi - \psi ||_{L^\infty_\rho([0,T] \times \mathbb{R}^3)}   \nonumber \\
  & =&  C_1 C^2\,  e^{2\rho l}  \left[ {1\over \rho^2} [1 - e^{-\rho t}] - {t\over \rho} e^{-\rho t} \right] 
  \, || \phi - \psi ||_{L^\infty_\rho([0,T] \times \mathbb{R}^3)} \nonumber  \\
  & \leq & C_1 C^2\, { e^{2\rho l} \over  \rho^2}  \, || \phi - \psi ||_{L^\infty_\rho([0,T] \times \mathbb{R}^3)}
\; .  \label{A2ineq7} 
\eea
(\ref{A2ineq7}) holds for any $T> T_0$. 
Take, say, $\rho = 1/2l$, and the scale $l$ and/or couplings in $V$ sufficiently small, so that 
$C_1 C^2\,  (e^{2\rho l} / \rho^2)   < 1$. Then  (\ref{A2ineq7}) shows that $A$ is a contraction. 
It follows that a unique solution to the integral equation (\ref{ie2})  satisfying the initial data on $ [-t_0, 0]$ exists on any interval $[0,T]$ with $T > T_0$.

\subsubsection{Stricly nonlocal interactions }
The case of strictly nonlocal interactions differs in basic ways. For such interactions the future delay is of infinite extent and each point inside a given domain $D$ (Fig. \ref{qlF2}) contributes to it. Hence, asymptotic switching off of interactions no longer makes  the argument leading to (\ref{A2ineq7}) possible.  
The demonstration of local existence of solutions still formally goes through since one may always specify an extension of the mapping $A$ over the infinite extent delay region. Existence of an infinity of solutions can thus be established. The causal effects, however, are no longer confined in a (small) finite region and cannot, in principle, be masked at larger scales as in the quasi-local case. 
Still, strictly nonlocal interactions may approximately behave as quasi-local ones if their delocalization  kernel falls off sufficiently rapidly beyond  a characteristic scale $\ell$, the long-range causal tails leaking out at long distances producing extremely small effects.

The classical IVP, though it serves well to demonstrate how nonlocality entails causality problems, is ultimately not of prime physical relevance. For this we have to turn to the quantum theory where 
one sums over configurations not restricted to solutions of the classical equations of 
motion.

\section{Evolution via time-blocked Hamiltonian \label{tbH}}
\setcounter{equation}{0}
\setcounter{Roman}{0} 
Though in this paper we adopt the Lagrangian formulation as the proper framework for both classical and  quantum treatment of nonlocal field theories,  we digress here to discuss a 
Hamiltonian formulation. Given our Lagrangian (\ref{act2}) one may 
proceed, as usual, to define a canonical momentum 
\beq 
\pi(\bfx,t) = {\partial {\cal L} \over \partial \dot{\phi}(\bfx,t) } = \dot{\phi}((\bfx,t)  \label{canmom}
\eeq
and the corresponding ``Hamiltonian" 
\beq H(t) = \int d^{(d-1)}x 
\left[ {1\over 2} \pi(\bfx,t)^2 + {1\over 2}(\nabla \phi(\bfx,t))^2 + {1\over 2} m^2 \phi(\bfx,t)^2 
+ V(\tlphi(\bfx,t))  \right] \, .   \label{H1}
\eeq
This Hamiltonian is positive. Through the $\tlphi$ dependence in the potential, however, it involves field functional dependence over times other than the time $t$ at which it is defined. Hence, as it is easily seen, the corresponding canonical equations fail to reproduce the correct equation of motion (\ref{eom1}).  The usual Hamiltonian formalism cannot be implemented in the presence of nonlocal interactions. 

Our results in the previous section, however, suggest that one should, instead, define a {\it smeared} Hamiltonian over time intervals longer than those in the delocalization kernel $F$. 
In this section we only consider kernels of compact support. 
Let $U(t)$ denote a  $C^\infty_c(\mathbb{R})$ `bump'  function such that $0\leq U \leq 1$ and 
\beq
U(t) = \left\{ \begin{array}{l l } 
1 & \mbox{for} \quad  |t| \leq 2\ell-\epsilon \\
0 &  \mbox{for} \quad  |t| \geq 2\ell+\epsilon
\end{array} \right.   \, . \label{pu1}
\eeq  
We now define a smeared Hamiltonian 
\beq 
 \tlH(t) =   \int dz^0 \, U(t-z^0) \, H(z^0)    \label{H2}
\eeq
which, by (\ref{pu1}), averages over an interval of length $2(2\ell +\epsilon)$ centered at $t$. In terms of this Hamiltonian one now has 
\beq 
{\delta \tlH \over \delta \pi(\bfx,t) }  = U(0) \pi(\bfx,t) = \pi(\bfx,t)   \label{ceom1}
\eeq
and 
\bea
{\delta  \tlH \over \delta \phi(\bfx,t) } & = & U(0) \Big[ - \nabla^2 \phi(\bfx,t) + m^2 \phi(\bfx,t) \Big]   \nonumber \\
&  &  \qquad + \int dz^0 U(t-z^0) \int dz^3 F(\bfx-\bfz, t-z^0) {\partial V(\tlphi(\bfz, z^0)) \over \partial \tlphi(z)} \nonumber \\
& = & \Big[ - \nabla^2 \phi(\bfx,t) + m^2 \phi(\bfx,t) \Big] 
+ \int dz^0 \int dz^3 F(x-z) {\partial V(\tlphi(z)) \over \partial \tlphi(z)}  \, , \label{ceom2}
\eea  
where in the second equality we used the fact that the compact support of $F$ enforces  $|t-z^0|\leq \ell$, and $U(w)=1$ for $|w|\leq\ell$. The canonical equations for $\tlH$ then 
\beq 
{\delta \tlH \over \delta \pi(\bfx,t) } =  \dot{\phi}(\bfx,t) \, , \qquad {\delta \tlH \over \delta \phi(\bfx,t) }=  - \dot{\pi}(\bfx,t)   \label{ceom3}) 
\eeq
reproduce the equation of motion (\ref{eom1}). 

We may then split the time axis into segments of size $4\ell$ and define a blocked Hamiltonian (\ref{H2}) on each segment. 
Take the cover of the time axis given by the union of the intervals 
\[ I_k = \big(4\ell(k - {1\over 2})-\epsilon,\,  4\ell(k+ {1\over 2})+ \epsilon\big) \, , \quad k=0, \pm1, \pm2, \ldots   \, ,  \]
$I_k$ being centered at $t_k\equiv 4\ell k$. 
Defining  $U_k (t) = U(t_k-t )$, and    
\beq 
u_k (t)= {U_k(t) \over \sum_l U_l(t)}   \, ,  \label{pu2} 
\eeq
one has the corresponding partition of unity  on $\mathbb{R}$:
\beq 1= \sum_k u_k(t)  \, .  \label{pu3}
\eeq
Note that, by (\ref{pu1}), for any given $t$, at most two terms can be non-zero in the sum in the denominator in (\ref{pu2}); and at most two terms can be non-zero in the sum in (\ref{pu3}).

We now define the blocked Hamiltonian on $I_k\times \mathbb{R}^3$ 
\beq 
\tlH(t_k) \equiv  \tlH_k = \int dz^0 \, u_k(z^0) \, H(z^0)  \, ,   \label{H2a}
\eeq
the canonical equations for which, as seen above, give the correct equations of motion at time $t_k$. 
The system may thus be described as evolving over successive time blocks via the blocked Hamiltonians $\tlH_k$ in a description that does not probe scales of order or smaller than the nonlocality  scale $\ell$. The description is invariant under shifts in the choice of times $t_k$ since any partition of unity (\ref{pu3}) is. This amounts to a quasilocal conservation statement for the blocked Hamiltonians in the following sense. Consider 
\bea
\dot{\tlH}(t) & = &   
  -  \int dz^0 \, {d\over dz^0}  u(t - z^0) \, H(z^0)  
 =   \int dz^0 \,  u(t - z^0) \, \dot{H}(z^0)   \nonumber \\
& = & \int dz^0 d^3{\bfz}\, \,  u(t-z^0) \Bigg[\Big[\ddot{\phi}(\bfz,z^0)  -[\nabla\phi(\bfz,z^0)]^2 + m^2 \phi(\bfz,z^0)\Big] \, \dot{\phi}(\bfz,z^0)  
+ V^\prime(\tlphi(\bfz,z^0))\, \dot{\tlphi}(\bfz,z^0) \Bigg]  \nonumber \\
& = & \int dz^0 d^3\bfz \,  \Bigg\{ \, u(t- z^0) \Big[ \ddot{\phi}(\bfz,z^0) -[\nabla\phi(\bfz,z^0)]^2 + m^2 \phi(\bfz,z^0)       \Big]     \nonumber \\
& & \quad + \int dy^0 d^3\bfy \;  u(t - y^0) V^\prime(\tlphi(\bfy, y^0) ) F(\bfy - \bfz, y^0-z^0) \Bigg\} 
\dot{\phi}(\bfz,z^0)     \, . \label{H3}
\eea
For a given $k$ consider $z^0$ such that $u(t_k- z^0)\not= 0$, i.e. $z^0\in I_k$. 
In the integrand in the last line in (\ref{H3}), the bounded support of kernel $F$ implies $|y^0- z^0| \leq \ell$.  
It follows  that  
\[ \sum_{l=k, k\pm 1} u(t_l -z^0) = \sum_{l=k, k\pm1} u(t_l -y^0) = 1 \, , \]  
where, in fact, only at most two out of the three terms in these sums can be nonzero. 
For any given $k$ then (\ref{H3}) gives 
\beq
\dot{\tlH}_{k-1} + \dot{\tlH}_k + \dot{\tlH}_{k+1} \equiv \sum_{j= k, k\pm1} {d\over d\kappa}\tlH(t_j + \kappa)\mid_{\kappa=0}  = 0 \, . \label{H4} 
\eeq  
by the equations of motion (\ref{eom1}).  
We thus have quasilocal conservation of blocked Hamiltonians, i.e., conservation involving   
a blocked Hamiltonian and at most one of its neighbors contributing in the overlap or ``spill-over" regions between neighboring block intervals $I_k$.

\section{The quantum theory  \label{QT}}
\setcounter{equation}{0}
\setcounter{Roman}{0} 
Quantization of (\ref{act1}) is straightforwardly performed via the path integral. The kinetic energy term, being of the standard local form, gives the usual scalar propagator 
\beq
\Delta(x-y) =  \int d^4k \, \Delta(k)\, e^{-ik(x-y)} = \int  {d^4 k \over (2\pi)^4}   \, {i  \over (k^2 - m^2 +i\epsilon) } \, e^{-ik(x-y)} \,  . \label{xprop1}
\eeq
An $n$-point interaction term $g_n\tlphi^n$  in $V(\tlphi)$ gives a nonlocal vertex factor 
$-i g_n \prod_{i=1}^n  F(x-x_i)$, and thus contributes a kernel factor $F(x-x_i)$ to a propagator attached to point $x_i$. 
Hence the propagator $\Delta(x-y)$ in each internal line in a graph is effectively replaced by a propagator: 
\beq 
\tlDl(x-y) = \int  d^4 u \,d^4 v\, F(x-u) \Delta(u-v) F(v-y) \, .  \label{tlxprop1}
\eeq
An equivalent set of rules is then given by local vertex factors $-ig_n$ and propagators (\ref{tlxprop1}):
\beq 
\tlDl(x-y) =  \int d^4k \, \tlDl(k)\, e^{-ik(x-y)} = \int  {d^4 k \over (2\pi)^4}   \, {i\, \hat{F}(k)^2 \over (k^2 - m^2 +i\epsilon) } \, e^{-ik(x-y)} \,  . \label{tlxprop1a}
\eeq

The amplitude for the general Feynman graph with $I$ internal lines, $L$ loops 
and $V$ vertices then  takes the form
\bea
A(\{p_j\})  & =  &   G(-i)^V\int \prod_{i=1}^L \, d^4k_i \, \prod_{j=1}^I \, \tlDl(q_j)    \nonumber \\
  & = & G(-i)^V {i^n\over (2\pi)^{4n}} \int \prod_{i=1}^L \, d^4k_i \, \prod_{j=1}^I \, {\hat{F}(q_j)^2 \over (q_j^2 - m^2 +i\epsilon) } \;   ,   \label{ampl} 
\eea
where $k_i$ denote $L$ independent loop momenta, $q_j$ denote the momenta for the internal lines, each $q_j$ being a linear combination of the $k_i$ and the external momenta $p_j$, and $G$ is the product of the coupling constant factors from the vertices.

\subsection{UV behavior and unitarity} 

Assume interactions of the form $\phi^n$ in $V$ in (\ref{act1}) - (\ref{act2}). 
All amplitudes (\ref{ampl}) are then UV finite. Indeed, before the introduction of delocalization kernels 
such interactions produce power divergences of superficial degree of divergence $4-E + (n-4)V_n$, where $E$ is the number of external legs and $V_n$ the number of vertices in the graph. 
The introduction of the delocalization kernel resulting in the $\hat{F}^2$ factor for each internal line in (\ref{ampl}) removes any such power divergences by its rapid decay property (\ref{rdf1}) (property I 
of admissible kernels, section 2). The same applies to a range of other interactions $V(\phi)$, in particular,  transcendental potentials admitting convergent power series  expansions.\footnote{ As noted before, interactions ensuring manifest Lipschitz-continuity in the classical theory (cf. section \ref{IVP-nl} on the classical IVP) would be examples of interest here. \label{FLcinter}}

The other property required of admissible kernels is that $\hat{F}(k)$ is an entire function of $k$ (Property II of section 2). This ensures that the Landau equations for locating the singularities of any given amplitude (\ref{ampl}) are not changed by the presence of $\hat{F}$ factors in the integrand. 
Their derivation \cite{ELOP}  is the same whether $\hat{F}$ is a polynomial entire function as in local theories, or a transcendental entire function as in the nonlocal theory case. 
Similarly, the derivation of the Cutkosky discontinuity (cutting) rules \cite{C}, \cite{ELOP} is unaffected since it only assumes that any $\hat{F}$ factors in the integrand in (\ref{ampl}) are entire functions of their arguments. It follows that, at least order by order in the perturbative expansion, the theory is unitary.

An important difference between polynomial and transcendendal $\hat{F}$ is that in the latter case continuation from Euclidean to Minkowski momenta by Wick rotation is generally no longer possible. 
It is not generally possible to `close the contour at infinity' due to the different growth behavior of transcendental entire functions in different directions in the complex plane. 
Note, again, in this connection that the rapid decay property is required to hold for all real values of the arguments, and so for both Minkowski and Euclidean momenta configurations.


\subsection{Causality}

As it is well known any formulation of causality in relativistic quantum field theory involving causal relations between spacetime events necessarily suffers from the fact that spacetime points cannot be 
pinpointed by wave packets built from physical, i.e. on-shell particles. Conditions can then be generally stated only in terms of some Green's or correlation functions. 
The relevant basic property is 
the decomposition of the causal (Feynman) propagator (\ref{xprop1}) into positive and negative frequency parts: 
\beq 
\Delta (x) = \theta(x^0) \Delta^+(x) + \theta(-x^0) \Delta^-(x)    \label{xprop2}
\eeq
with 
\beq 
\Delta^{\pm}(x) =  \int {d^4k \over (2\pi)^3 }\, \theta(\pm k_0)\,  \delta(k^2 - m^2)\, e^{-ik x} \equiv \int d^4 k \,  \Delta^{\pm}(k)  \,  e^{-ik x}     \,.    \label{xprop3}
\eeq 

When the basic structure (\ref{xprop1}) is modified as  in (\ref{tlxprop1a}), 
this decomposition no longer holds in the form (\ref{xprop2}). 
Inserting (\ref{xprop2}) in (\ref{tlxprop1})  one obtains
\bea  
\mbox{\hspace{-1cm}}\tlDl(x-y) &  = &  \int d^4 u\, d^4 v F(x-u) F(v-y)   \nonumber \\
   & & \mbox{\hspace{1cm}}  \cdot [ \theta(u^0 - v^0) \Delta^+(u-v) + \theta(v^0-u^0) \Delta^-(u-v) ] 
 \label{tlxprop2}  \\
& = & \int d^4 q e^{-iq(x-y)} \int {d\omega \over 2\pi} \, {i\over \omega + i\epsilon}  \nonumber \\
& &\mbox{\hspace{-1.2cm}} \cdot \left\{ e^{-i\omega (x^0 - y^0 )} \hat{F}^2({\bf q}, q_0 +\omega) \Delta^+({\bf q}, q_0) 
  + e^{-i \omega(y^0 - x^0 )}\hat{F}^2({\bf q}, q_0 -\omega) \Delta^-({\bf q}, q_0)   \right\} .
   \label{tlxprop3}
  \eea
 Expanding the entire functions $\hat{F}^2({\bf q}, q_0 \pm\omega)$ in an $\omega$-power series (\ref{tlxprop3}) can be written in the form 
 \beq 
\tlDl(x-y) =    \tlDl_c (x-y)  + \tlDl_{nc}(x-y)     \,. \label{tlxprop4}
\eeq 
Here $\tlDl_c$ is a causal propagator defined by 
\beq 
  \tlDl_c (x) =  \theta(x^0) \tlDl^+(x) + \theta(-x^0) \tlDl^-(x)   \, ,   \label{tlxprop5} 
\eeq
where 
\beq 
\tlDl^{\pm}(x)  
\equiv \int d^4 q \, \hat{F}^2 (q)  \Delta^{\pm}(q)  \,  e^{-iq x}   = \int {d^3q \over (2\pi)^3 }{1\over 2\omega_q} \hat{F}^2 ({\bf q}, \pm \omega_q)\,  e^{i{\bf q}\cdot {\bf x}} e^{\mp i \omega_q x^0}       \label{tlxprop5a}
\eeq 
with $\omega_q = \sqrt{{\bf q}^2 + m^2}$. The remainder $\tlDl_{nc}$ is given by 
\beq 
\tlDl_{nc} (x-y) = i \sum_ {m\geq 1}   {i^{m-1}\over m!} \,\delta^{(m-1)}(x^0 - y^0) \left[ \tlDl^{+ \,(m)} (x-y) -  \tlDl^{- \,(m)} (x-y)  \right]  \, ,\label{tlxprop6}
\eeq
where   
\beq 
\tlDl^{\pm\, (m)}(x)  
\equiv \int d^4 q \, \hat{F}^{2\,(m)}(q)  \Delta^{\pm}(q)  \,  e^{-iq x}   = \int {d^3q \over (2\pi)^3 }{1\over 2\omega_q} \hat{F}^{2\, (m)}({\bf q}, \pm \omega_q)\,  e^{i{\bf q}\cdot {\bf x}} e^{\mp i \omega_q x^0}       \label{tlxprop6a}
\eeq 
with 
\beq
\hat{F}^{2\, (m)}(q) = \partial^m \hat{F}^2({\bf q}, q_0) /\partial q_0^m,  \quad m=1,2,\ldots  
\label{tlxprop6b}
\eeq

Some remarks concerning these formulas should be made. 
(\ref{tlxprop4}) - (\ref{tlxprop6b}) hold for any entire function $\hat{F}$. In particular, they hold for polynomial $\hat{F}$ which is in fact the familiar case of local theories with finite order derivative couplings and/or nonzero spin fields (modulo the appropriate tensor structures). In that case the 
sum in (\ref{tlxprop6}) is finite, i.e., $\tlDl_{nc}$  consists of a finite number of contact terms. 
These contact terms arise from transporting derivatives across the theta functions in (\ref{xprop2})
into the $\Delta^{\pm}$ functions (cf. (\ref{tlxprop5a})). Such contact terms are simply dropped since they can be absorbed into a {\it finite} number of  local counterterms as discussed in the 
textbooks.\footnote{This is the known ambiguity in the  definition of  the (free) Feynman  Green's function:  polynomial factors, due, for example, nonzero spin, can either be included in the definition of the $\Delta^{\pm}$s or  act on the entire scalar propagator (\ref{xprop2}); the latter convention is nearly universally adopted. The ambiguity amounts to (Lorentz non-invariant) local counterterms. The most complete and mathematically careful discussion 
is given in the classic textbook \cite{BS}, see also \cite{tHV1}. \label{Flct} } 

\begin{figure}[ht]
\begin{center}
\includegraphics[width=0.7\textwidth]{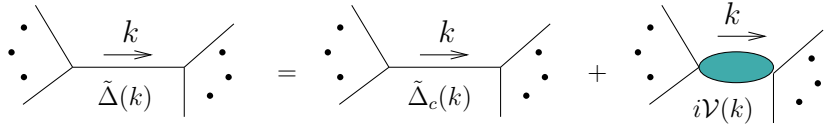}
\end{center}
\caption{The propagator $\tlDl(k)$ can be decomposed in the sum of a causal propagator $\tlDl_c(k)$ and $\tlDl_{nc}(k)$  which is entire analytic in the energy plane and thus can be treated as a nonlocal vertex  $i {\cal V}(k)$ for which momentum conservation holds as usual.
 \label{qlF3}}
\end{figure}

In the case of nonlocal interactions, however, the sum in (\ref{tlxprop6}) is an infinite sum that no longer can be removed by a finite number of local counterterms. The contact terms now sum up to a transcendental entire function, a nonlocal contribution that cannot be dropped: the non-polynomial asymptotic behavior renders the propagator $\tlDl$, given by  (\ref{tlxprop1a}), {\it not} equal to the causal propagator $\tlDl_c$ given by (\ref{tlxprop5}). By the same token a (appropriately subtracted) 
Kall\'en-Lehmann representation cannot be obtained for (\ref{tlxprop1a}). 
Note in this connection  that in the expression (\ref{tlxprop6}) for $\tlDl_{nc}$ the quantity 
\[ \tlDl^{+ \,(m)} (x-y) -  \tlDl^{- \,(m)} (x-y)  =  
 \int d^4 q \, \hat{F}^{2\,(m)}(q) [ \Delta^+(q)   - \Delta^-(q)]   \,  e^{-iq x}    \]
 involves the field commutator Fourier transform $  \Delta^+(q)   - \Delta^-(q)$, which, however, 
upon integration will not vanish for spacelike  distances due to the presence of the $\hat{F}^{2\,(m)}$ factors. In momentum space (\ref{tlxprop6}) is given by 
\bea
\mbox{\hspace{-1cm}} \tlDl_{nc}(k) & = &\int  d^4 x\, \tlDl_{nc}(x)\,  e ^{ikx} \nonumber \\
& = & i \sum_{m\geq 1} {1\over m!} {1\over 2\omega_k} \left[ \hat{F}^{2\,(m)}({\bf k}, \omega_k) \, (k_0 - \omega_k)^{m-1}  - \hat{F}^{2\,(m)}({\bf k}, -\omega_k)\, (k_0 + \omega_k)^{m-1} \right] \label{tlkprop1} \\
& \equiv & i {\cal V}(k) \nonumber 
\eea
as the convergent expansion in positive powers of an entire function. Thus, $\tlDl_{nc}$  manifestly does not contribute to absorptive parts and can in fact be viewed as a nonlocal vertex 
$ i {\cal V}(k)$ (Figure \ref{qlF3}). 

In the case of kernels of bounded support the following observation can now be made. 
If  $|x^0 - y^0| >  2\ell $  in (\ref{tlxprop2}), then  one necessarily has $\sign (u^0-v^0) = \sign (x^0 - y^0)$. Hence  
\bea
\tlDl (x-y) & = & \theta\big( |x^0 - y^0| -  2\ell\big)\, \tlDl(x-y) +  \theta\big(2\ell -  |x^0 - y^0| \big)\, \tlDl(x-y)  \nonumber \\
& = & \theta\big( |x^0 - y^0| -  2\ell\big)\,  \tlDl_c(x-y) +  \theta\big(2\ell -  |x^0 - y^0| \big)\, \tlDl(x-y)  \nonumber \\
& = & \tlDl_c(x-y)  +  \theta\big(2\ell -  |x^0 - y^0| \big)\, \big [ \tlDl(x-y)  - \tlDl_c(x-y) \big] . 
\label{tlxprop7}
\eea
Thus, in coordinate space, the difference between $\tlDl$  and the causal propagator, i.e. , $\tlDl_{nc} = \tlDl(x) - \tlDl_c$,  
arises entirely from inside the bounded region of the kernel support and its effect is confined in it. 
In momentum space, since for sufficiently small $\ell$, $\hat{F}(k)$ is essentially constant for all $k$ less than of order $1/\ell$, and is of rapid decay for $k > 1/ \ell$, it is evident from (\ref{tlkprop1}) that $\tlDl_{nc}(k)$ is appreciably nonvanishing only for momenta of order $1/\ell$. 
Note how the (classical) picture of evolution over time-blocks of size $\sim \ell$ in the section \ref{tbH}  
accords with these quantum propagation properties. 

Having examined the structure of the propagator we may now consider the conditions imposed on amplitudes by causality. 

\subsubsection{Local interaction} 
It will be useful to first recall the local case. The general causality condition is the  Bogoliubov causality condition \cite{BS}, \cite{Bo}.  It can be stated, in various slightly different but equivalent versions, in terms of amplitudes and thus in terms of diagrams. 
(Conditions on $n$-point Wightman functions under appropriate  permutations of their arguments, a special case of which is the frequently stated condition of commutativity of field operators at spacelike distances, reduce, when diagrammatically expressed,  to special versions of the Bogoliubov conditions.) 

In the local interaction case the propagators are given by (\ref{xprop1}), (\ref{xprop2}) and (\ref{xprop3}) with local vertex factors $-ig_n$.
Following \cite{V}, \cite{tHV1} we conveniently state the Bogolyubov causality condition in the form:

\parbox{14cm}
{\epsfysize=3cm \epsfxsize=13cm \epsfbox{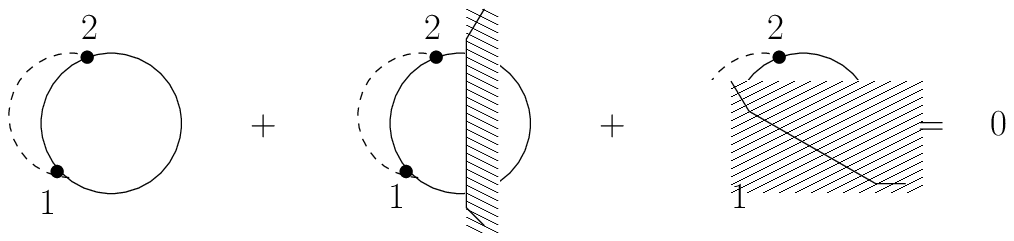}}     
\hfill \parbox{0.7cm}{\begin{equation}\label{bcc1}
\end{equation}}\\[0.7cm]

\begin{figure}[ht]
\begin{center}
\epsfysize=6.5cm\epsfbox{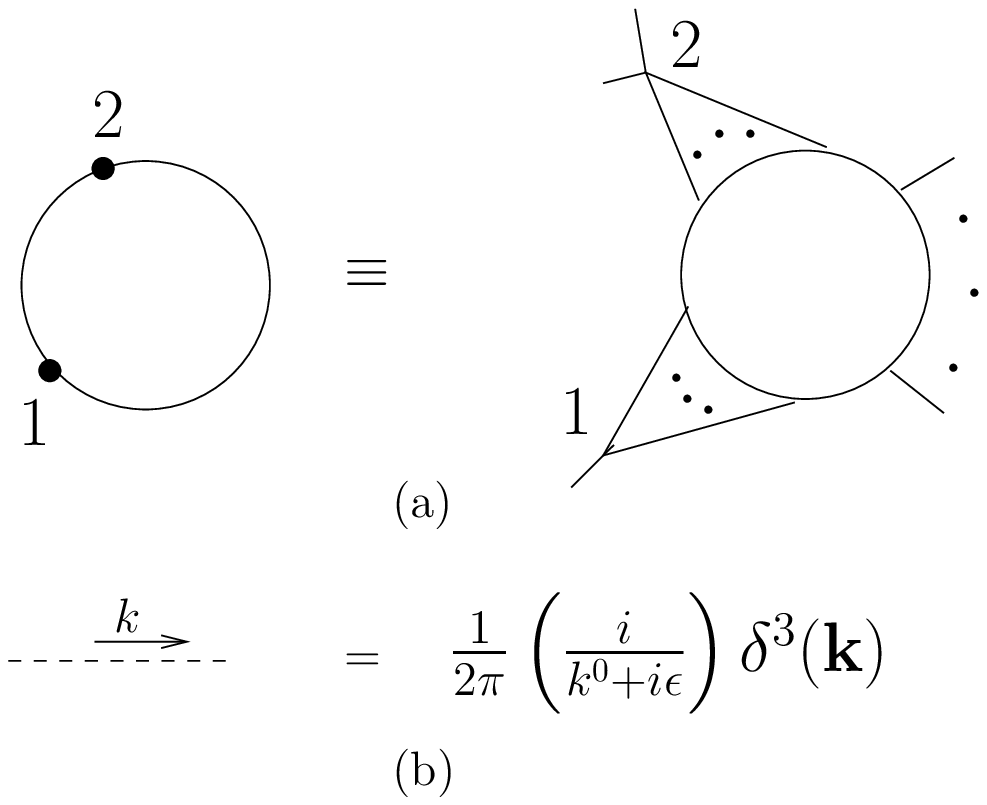}
\end{center}
\caption{Diagrammatic notation in (\ref{bcc1}) (cf.  text).    \label{qlF5}}
\end{figure}

\begin{figure}[ht]
\begin{center}
\epsfysize=4cm\epsfbox{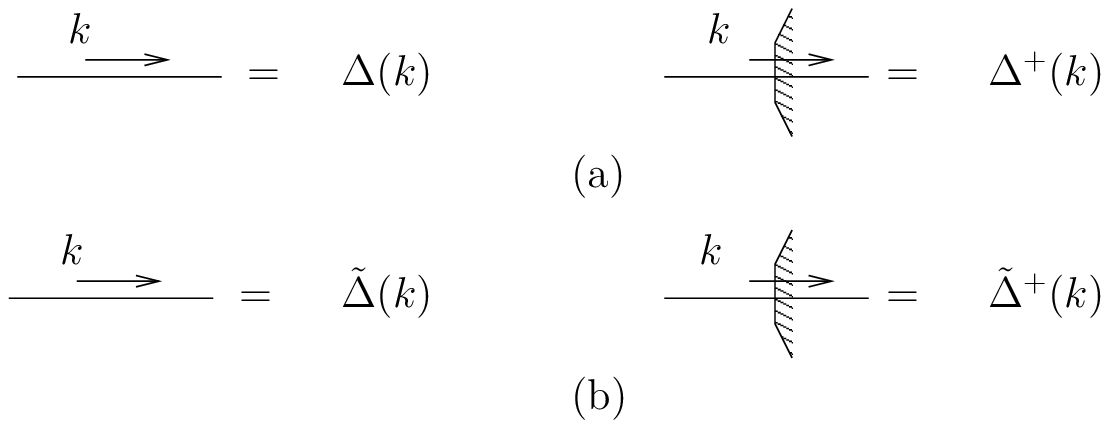}
\end{center}
\caption{Propagator and cut lines  in (a) local theory, eq. (\ref{bcc1}); (b) nonlocal theory, eq. (\ref{bcc2}).   \label{qlF6}}
\end{figure}

Here the blob represents any given diagram or collection of diagrams  with a given number of external legs. The external vertices, labeled 1 and 2, to which two or more of these legs are attached, have been selected and explicitly indicated; nothing else is depicted explicitly, see Fig. \ref{qlF5} (a). 
Vertices 1 and 2 may be any interaction vertices present in the Lagrangian or represent the insertion of operators used to probe the process. 
The broken line arises from the momentum space representation of theta functions introduced for time-ordering vertices 1 and 2 in the coordinate space 
statement of the condition \cite{BS}; it amounts to a (non-covariant) propagator connecting 
the two vertices given in Fig. \ref{qlF5} (b). 
The shaded lines in (\ref{bcc1}) represent the sum over all Cutkosky cuts placed in the manner indicated. Energy flows from the unshaded to the shaded side, each cut propagator being replaced by $\Delta^+(k)$ (Fig. \ref{qlF6}(a)). 

Because energy is conserved at vertices a region on the unshaded side must be connected to incoming lines, and a region on the shaded sided to outgoing lines. 
On the shaded side Feynman rules are those of $S^\dagger$, i.e., complex conjugated propagators and vertex factors. Thus, 
the second term represents the sum over all Cutkoski cuts with vertices 1 and 2 on the unshaded side; the sum in the third term is over cuts with vertex 2 on the shaded side and vertex 1 on the unshaded side. An efficient derivation proceeds from the Veltman largest time equation which employs the representation (\ref{xprop2}) leading to general cutting formulas for any (set of) graph(s) \cite{V}. (\ref{bcc1})) is then obtained as a particular application of these cutting formulas.\footnote{The unitarity equations expressing the absorptive parts as a sum over Cutkosky cuts can  also be obtained as another   application of these  cutting formulas. \label{Funiteqs}} The well-known details are given in \cite{V} , \cite{tHV1}.

\subsubsection{Nonlocal interaction} 
\begin{figure}[ht]
\begin{center}
\epsfysize=2.5cm\epsfxsize=15cm\epsfbox{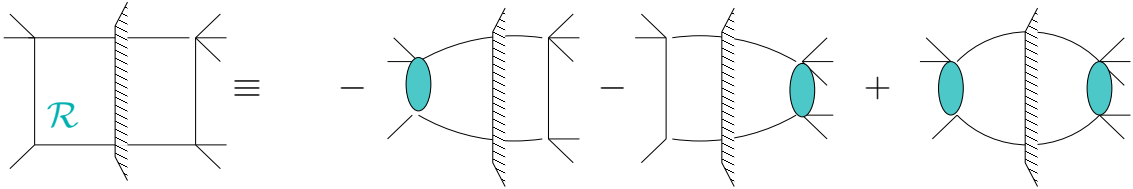}
\end{center}
\caption{The set of reduced graphs that arise from a cut box graph.   \label{qlF8}}
\end{figure}

In the presence of nonlocal interactions (\ref{xprop2}) is replaced by (\ref{tlxprop4}).  The presence of $\tlDl_{nc}$ means that the derivation leading to (\ref{bcc1}) no longer holds. To obtain an appropriate extension of (\ref{bcc1}) in the nonlocal case we proceed as follows. Start with the causal propagator given by (\ref{tlxprop5}) - (\ref{tlxprop5a}).  Using this causal propagator the derivation leading to (\ref{bcc1}) now applies, and one regains (\ref{bcc1}) but now  with propagators $\tlDl_c$ and cut lines $\tlDl^+$. (Incidentally, this would be the BCC equation for 
nonzero spin fields and/or local derivative interactions, i.e., cases with appropriate polynomial $F$, as discussed above.) We next substitute $\tlDl_c = \tlDl - \tlDl_{nc}$ for each propagator. 
In this manner we obtain:

\parbox{14cm}
{\epsfysize=7cm \epsfxsize=13cm \epsfbox{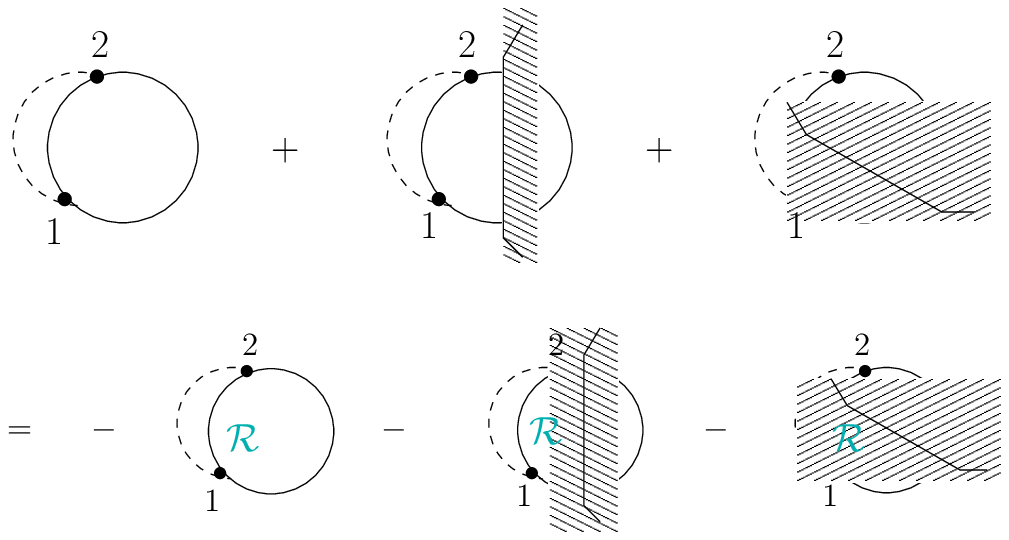}}     
\hfill \parbox{0.7cm}{\begin{equation}\label{bcc2}
\end{equation}}\\[0.7cm]
The l.h.s in (\ref{bcc2}) is as in (\ref{bcc1}) but now applied to the our nonlocal theory (\ref{act2}) - (\ref{deloc1}), i.e.  all propagators are given by $\tlDl(k)$ on the unshaded side and its complex conjugate on the shaded side, whereas all cut lines represent $\tlDl^+(k)$, cf. Fig. \ref{qlF6} (b). To each term on the l.h.s. there corresponds a set of reduced graphs, indicated by the insertion of the label ${\cal R}$, obtained by replacing its propagators (uncut internal lines) by the nonlocal vertex $-i{\cal V}(k)$ in all possible ways. 
 A simple example is given in Fig. \ref{qlF8}.
The r.h.s. in (\ref{bcc2}) then consists of summing over all such reduced graphs. 
(\ref{bcc2}) is our extension of the Bogoliubov causality equation in the presence of nonlocal interactions. The r.h.s. represents the corrections to the local theory condition (\ref{bcc1}) due to nonlocality.

In the case of quasi-local, i.e., bounded support  interactions these corrections are  generally 
extremely small since, as pointed above, a nonlocal vertex ${\cal V}(k)$ differs from zero essentially only around momenta of order $1/\ell$.  
Of particular relevance are processes involving scales longer than  the nonlocality scale $\ell$. 
Except for internal momenta circulating in loops, momenta flowing through the diagram are below $1/\ell$ and the corresponding internal lines gives essentially zero contribution to the nonlocal vertex (\ref{tlkprop1}). 
It is natural in such a case to consider, in particular, the two external vertices 1 and 2 used to probe causality to be connected to the rest of the diagram by tree branches carrying momenta well below $1/\ell$ into the diagram. This would represent measurements over spacetime scales (much) larger than $\ell$. 
Contributions from the nonlocal vertex in internal loops are small because of the rapid decay properties of $\hat{F}$  cutting off momenta above $1/\ell$ - cf. remark following (\ref{tlxprop7}). {\it The theory at scales longer than  $\ell$ behaves essentially as a local theory with a cutoff $1/\ell$}. This may be viewed as providing the quantum underpinning of our remarks in the classical context in the latter part of section \ref{IVP-nl}.

In the case of strictly nonlocal, i.e., unbounded support interactions characterized by some scale $\ell$ the noncausal effects can  ``leak" out of regions of size $\ell$ and be more or less pronounced 
depending on the form of the delocalization kernel $F$. If $F$ decays exponentially outside a region of size $\ell$, any non causal effects can be extremely small at large distances relative to $\ell$. 
Such nonlocal interactions would be hardly distinguishable from quasi-local ones, at least as long as 
small acausal effects leaking out do not accumulate in some special processes.

\section{Discussion and outlook \label{DO}}
In summary, we studied scalar field theories with interactions of delocalized fields, the delocalization being  specified through a nonlocal integral kernel $F(x-y)$. We imposed conditions on such kernels 
to insure UV finiteness and unitarity of amplitudes. 
Kernels satisfying such conditions are smooth functions of rapid decay and classified as either strictly nonlocal or quasi-local (bounded support) kernels. 
In the quantum theory this may be described as the kernels being chosen in the appropriate test function spaces that regularize the quantum fields (tempered distributions).  
Using this framework we gave a detailed treatment of the classical IVP. 
The introduction of nonlocal kernels results into partial integro-differential equations of motion and an IVP with past and future delays. We gave rigorous proofs of the existence but also the accompanying complete lack of uniqueness of solutions due to the future delays. This is, of course, the manifestation of acausality. 
We saw how the acausality effects are mitigated when confined in a region of limited extent. 
Passing to the quantum theory we derived a generalization of the equation for the Bogolyubov causality condition on amplitudes. This generalization, eq. (\ref{bcc2}), explicitly shows how the terms in the equation expressing causal propagation must be supplemented to include the effects of nonlocality.  
 
As discussed in section \ref{QT} the structure of these acausal corrections is such that, for quasi-local kernels of length size $\ell$, their effect is confined within regions of size $\ell$. This is no longer true for strictly nonlocal kernels where the acausal effects leak out. But they can be very small for 
sufficiently (exponentially) fast decaying kernels. 
In this connection one may recall the example of the Lee-Wick prescription which, though unrelated to the nonlocal theories studied here,  also entails 
acausal tails but generally falling off only as powers of the distance. They, nonetheless, apparently produce tiny  effects \cite{Oetal}, \cite{TT1}. For strictly nonlocal interactions, however, the possibility always remains of devising certain processes 
where small acausal effects leaking  out to longer distances may accumulate. 

In any event our results indicate that quasi-local interactions are the most appealing. They  
possess the good UV behavior conferred by nonlocality  while they mitigate acausal effects by 
actually confining them within the bounded support region of the kernels given by some characteristic scale $\ell$. In fact, as we saw in the last section, for physics at momenta below 
a cutoff of order $1/\ell$ they essentially behave as local theories. Microscopic acausalities confined within a small length scale $\ell$ are not necessarily bad, and may even be desirable for certain applications, in particular, in the very early universe \cite{TT1}.

We dealt exclusively with scalar theories in this paper. Coupling to fermions would not appear to present any problems. Extension to include gauge interactions, however, is not 
straightforward. This is because gauge invariance relates the interaction and the free parts  of the action. Thus, they no longer can be independently modified without breaking the invariance. In particular, one can not delocalize just the interactions.  One approach is to delocalize gauge invariant or covariant field combinations such as $F_{\mu\nu}$ or $R_{\mu\nu}$. This is the approach followed in \cite{TT}, also \cite{BCKM}, \cite{Mod1}. It typically results in superrenormalizability rather than complete UV finiteness, as was already realized in the early work on nonlocal QED \cite{F}.   Another approach \cite{Eet} is to first do modify the interactions, thus breaking gauge invariance, and then try to restore it by the iterative addition of an infinite series of new interaction terms to the action. 
Such constructions, though, are rather unwieldy and of doubtful action convergence properties, and have not been completely carried out explicitly.  
These, however, are not the only possible approaches for introducing nonlocality in gauge theories. Gravity, in particular, is special in the sense that the basic fields are tensors rather than connections, and this offers some additional possibilities, which we will consider elsewhere. 

Returning to scalar fields, one may remark that, at least in the absence of supersymmetry, they are difficult to accommodate within the usual framework of local field theory, i.e., incorporate 
as non-trivial (non-perturbatively existing) theories. 
In that they may be somewhat similar to gravity (the other even-spin bosonic field). 
There is always an inherent, wide arbitrariness in the potential of scalar fields that may be only resolved by the presence of a non-Gaussian (Wilson-Fisher type) UV fixed point assuring existence. Nonlocality, which, as expounded above, can accommodate general potentials, may offer  another way to accomplish this. It is in fact amusing that quasi-local interactions to some degree 
mimic the presence of an UV fixed point, as will be discussed elsewhere. 

Finally, one might consider the deeper question of how delocalized fields such as those studied here may arise. In this paper we introduced a scalar density function $F(x)$  on the spacetime manifold whose role is to delocalize fields. We then investigated the properties $F$ must possess and 
the behavior of the resulting theories of interacting delocalized fields in some detail.  
We did not inquire as to any possible physical basis or mechanisms underlying the presence of $F$. 
Addressing this question would require a deeper theory.

\end{document}